\documentclass[aps,english,showpacs,twocolumn,pra]{revtex4-1}
\usepackage{amsfonts}
\usepackage{amssymb}
\usepackage{amsmath}
\usepackage{graphicx}
\usepackage{epsfig}

\begin{document}

\title{Flat band induced by the interplay of synthetic magnetic flux and
non-Hermiticity}
\author{L. Jin}
\email{jinliang@nankai.edu.cn}
\affiliation{School of Physics, Nankai University, Tianjin 300071, China}

\begin{abstract}
A flat band is nondispersive and formed under destructive interference.
Although flat bands are found in various Hermitian systems, to realize a
flat band in non-Hermitian systems is an interesting task. Here, we propose
a flat band in a parity-time symmetric non-Hermitian lattice. The proposed
flat band has entirely real energy and is formed at an appropriate match
between synthetic magnetic flux and non-Hermiticity. The flat band energy is
tunable. At a weak intercell coupling, the flat band is isolated; whereas at
a strong intercell coupling, it intersects with the dispersive band at the
non-Hermitian phase transition point. The eigenstates of the flat band are
compact localized states and are confined in one, two, or three unit cells
at the edges or inside the non-Hermitian lattice.
\end{abstract}

\maketitle

\section{Introduction}

A flat band is entirely constituted by degenerate states, is nondispersive,
and has a zero group velocity. The superposition of nondispersive flat band
modes propagates without diffraction. A flat band exhibits many peculiar
features of wave propagation~\cite{LiebNJP}, localization length scaling~%
\cite{Scaling,LGeLL}, and unconventional Anderson localization~\cite{Baboux}%
. A simple model with a flat band is the Lieb lattice: A zero-energy flat
band intersects two linearly dispersive bands at the Dirac point~\cite{Lieb}%
. The Lieb lattice is a line-centered square lattice that can be realized in
photonics by trapping ions in optical lattices~\cite{VA} or through the
laser writing technology in optical waveguides~\cite%
{MolinaPRL,ThomsonPRL2015,Xia}. In addition, optical lattices in other
geometric structures of diamond (rhombic) \cite{CEC,Bermudez,Scaling},
kagome \cite{Yamamoto,Ventra}, honeycomb \cite{Wu,Jacqmin}, pyrochlore \cite%
{Shukla10,Shukla18}, and dice ($\mathcal{T}_{3}$) \cite{Vidal,Kolovsky}
lattices support flat bands \cite{FlachPRB17,LeykamAPX}.

A magnetic flux renders flat bands without disorders; thus, particles and
light can be perfectly trapped in a compact localization, following the
trapping mechanism of destructive interference. This mechanism differs from
the trapping mechanism in a spatially localized state induced by disorders~%
\cite{Anderson}. It has been proposed that the Aharonov-Bohm (AB) cage is
induced by magnetic flux in a two-dimensional dice lattice~\cite{Vidal} and
a quasi-one-dimensional (quasi-1D) diamond lattice~\cite{JC,CEC,SLOL}. The
AB cage has been experimentally observed in quasi-1D coupled waveguide
lattices \cite{OLMukherjee,PRL18,ASarXiv}.

Parity-time ($\mathcal{PT}$) symmetric non-Hermitian systems are manifested
in various microwave, optical, acoustic, and electronic systems~\cite%
{Bender98,NM,Bendix,LJin,Joglekar,Schomerus,Zeuner,SHFan,Dembowski,Ruschhaupt,El,AGuo,CE,CHLee,HJing,BP,LChang,Zhu,Alu,BHe,ZZhang,KDing,PTRevLGe,Kominis,PTRevAlu,Longhi18,PTRev}%
. They have potential applications in nonreciprocal dynamics~\cite%
{CPA,LFeng,BHePRL,LJinPRL}, topological energy transfer~\cite%
{Xu,Assawaworrarit}, and novel lasing~\cite%
{LFengScience,Hodaei,YangPNAS,Harari}. Recently, investigations on flat
bands have been extended to $\mathcal{PT}$-symmetric non-Hermitian systems~%
\cite{Molina,LGeLieb,RamezaniFB,LeykamFB,LGePR,LGePRL2018}. It is shown that
the Lieb stripe maintains a flat band regardless of the non-Hermiticity of $%
\mathcal{PT}$-symmetric gain and loss~\cite{Molina}. $\mathcal{PT}$%
-symmetric gain and loss perturbation lifts the degeneracy of a flat band in
a stub lattice~\cite{LGeLieb}. A flat band can be formed by non-Hermiticity
at the $\mathcal{PT}$-symmetric phase transition point in a triangular
lattice~\cite{RamezaniFB} and a cross-stitch lattice~\cite{LGePR}, and a
polynomial increase of the intensity for a single site excitation has been
revealed~\cite{LeykamFB,LGePR}. In additon, another type of zero-energy flat
band with net amplification and attenuation emerging from lattices under
non-Hermitian particle-hole symmetry has been investigated~\cite{LGePRL2018}.

In this paper, we report a new configuration of a quasi-1D $\mathcal{PT}$%
-symmetric non-Hermitian lattice. The lattice unit cell comprises a $%
\mathcal{PT}$-symmetric non-Hermitian triangular ring enclosed synthetic
magnetic flux, which is cross-stitch-coupled along the translationally
invariant direction. A real-energy flat band forms under destructive
interference at an appropriate match between synthetic magnetic flux and
non-Hermiticity. The energy of the flat band is enforced to be zero by
chiral symmetry; the intracell coupling between the active and dissipative
lattices breaks chiral symmetry, thus enabling tuning of the flat band
energy. The flat band can be isolated or intersected with dispersive bands,
where the band intersections are exceptional points (EPs). At EPs, the
degree of non-Hermiticity corresponds to the $\mathcal{PT}$-symmetric phase
transition threshold of the lattice. The flat band consists of compact
localized states (CLSs), which at the band intersections are the bound
states in the continuum. CLSs are confined in the single unit cell of the
system in the absence of chiral symmetry; by contrast, zero-energy flat band
CLSs can distribute in two or three unit cells.

The paper is organized as follows. In Sec. \ref{II}, a quasi-1D $\mathcal{PT}
$-symmetric non-Hermitian system is modeled. In Sec. \ref{III}, the energy
band structure of the non-Hermitian system is studied. The conditions for
forming a flat band, $\mathcal{PT}$-symmetric phase transition, and energy
band structures are also explored. In Sec. \ref{IV}, the localized modes and
edge modes of the flat band are shown. In Sec. \ref{V}, the results are
summarized and the prospects are presented.

\section{$\mathcal{PT}$-symmetric non-Hermitian lattice}

\label{II} 
\begin{figure}[t]
\includegraphics[ bb=0 0 340 340, width=8.8 cm, clip]{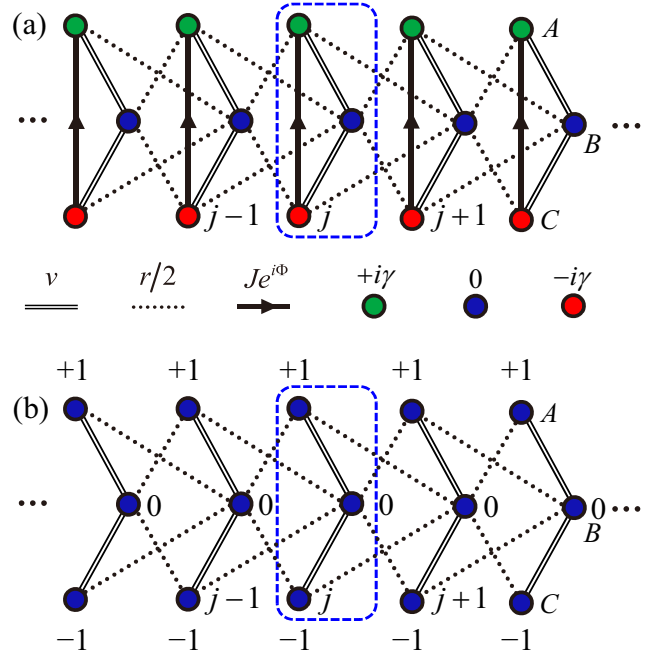}
\caption{Schematics of (a) a $\mathcal{PT}$-symmetric non-Hermitian lattice and (b) a Hermitian cross-stitch lattice. Sublattice $A$ ($C$) with
gain (loss) is depicted in green (red), whereas the passive sublattice ($B$) is depicted in blue. The
coupling between sublattices $A$ and $C$ is asymmetric; the
arrows denote the phase direction. The dashed blue rectangle indicates the lattice unit cell.
The dotted black lines represent the cross-stitch intercell coupling between the sublattices. A compact localized state of antisymmetric excitation $(1,0,-1)$ is shown in (b).
} \label{fig1}
\end{figure}

The non-Hermitian lattice considered in this study is
schematically illustrated in Fig.~\ref{fig1}(a). Three types of sublattices $%
A$, $B$, and $C$ are active, passive, and dissipative, respectively; the
gain and loss rates are $\gamma $~\cite{AGuo,CE,BP}. $A_{j}$, $B_{j}$, and $%
C_{j}$ constitute the $j$-th unit cell (denoted by the dashed blue
rectangle). The system is $\mathcal{PT} $-symmetric with respect to the
parity operation $\mathcal{P}A_{j}\mathcal{P}^{-1}=C_{N+1-j}$, $\mathcal{P}%
B_{j}\mathcal{P}^{-1}=B_{N+1-j}$ and the time-reversal operation $\mathcal{T}%
i\mathcal{T}^{-1}=-i$. $N$ is the total site number of each sublattice.

The tight-binding Hamiltonian is given by
\begin{eqnarray}
H &=&\sum_{j=1}^{N}[\frac{r}{2}(a_{j}^{\dagger }b_{j-1}+a_{j}^{\dagger
}b_{j+1}+c_{j}^{\dagger }b_{j-1}+c_{j}^{\dagger }b_{j+1})+Je^{i\Phi
}a_{j}^{\dagger }c_{j}  \notag \\
&&+v(c_{j}^{\dagger }b_{j}+b_{j}^{\dagger }a_{j})+\mathrm{H.c.}]+i\gamma
(a_{j}^{\dagger }a_{j}-c_{j}^{\dagger }c_{j}),
\end{eqnarray}%
where $a_{j}^{\dagger }$ ($a_{j}$), $b_{j}^{\dagger }$ ($b_{j}$), and $%
c_{j}^{\dagger }$ ($c_{j}$) are the creation (annihilation) operators that
satisfy the periodical boundary condition $a_{N+1}=a_{1}$, $b_{N+1}=b_{1}$,
and $c_{N+1}=c_{1}$, respectively. The system parameters ($\gamma $, $v$, $J$%
, $r$) are all considered to be positive real values without loss of
generality. The cross-stitch intercell coupling strength is $r/2$ ($r\neq 0$%
) \cite{SFlachPRL113,SFlachEPL,Nori}. The passive sublattice couples to the
active and dissipative sublattices with strength $v$. The asymmetric
coupling with the nonreciprocal phase factor $Je^{\pm i\Phi }$ between the
active and dissipative sublattices induces a gauge invariant synthetic
magnetic flux in each triangular unit cell. A nonreciprocal phase factor can
be realized in optical systems through chiral-light interaction~\cite{Ramos}%
, dynamical modulation~\cite{Yu}, photon-phonon interaction~\cite{ELi}, and
optical path length imbalance~\cite{Hafezi} in coupled waveguides and
resonators. This factor can also be realized through laser-induced tunneling~%
\cite{Bloch} and shaking of the lattice or Floquet engineering~\cite{Zhai}
for cold atoms in optical lattices \cite{Goldman16}. Moreover, synthetic
magnetic flux has been realized beyond optics in microwave~\cite{Sliwa} and
acoustic~\cite{Alu14} regimes.

By applying a Fourier transformation $a_{k}=N^{-1/2}%
\sum_{j=1}^{N}e^{ikj}a_{j}$, $b_{k}=N^{-1/2}\sum_{j=1}^{N}e^{ikj}b_{j}$, $%
c_{k}=N^{-1/2}\sum_{j=1}^{N}e^{ikj}c_{j}$, where the wave vector $k=2\pi n/N$
(integer $n\in \lbrack 1,N]$) and the Hamiltonian in the momentum space is
rewritten as $H=\sum_{k}H_{k}$, where $H_{k}$ is given by
\begin{equation}
H_{k}=\left(
\begin{array}{ccc}
i\gamma & v+r\cos k & Je^{i\Phi } \\
v+r\cos k & 0 & v+r\cos k \\
Je^{-i\Phi } & v+r\cos k & -i\gamma%
\end{array}%
\right) .  \label{Hk}
\end{equation}%
The $3\times 3$ matrix $H_{k}$ is $\mathcal{PT}$-symmetric under
the definition of the parity operator
\begin{equation}
\mathcal{P}=\left(
\begin{array}{ccc}
0 & 0 & 1 \\
0 & 1 & 0 \\
1 & 0 & 0%
\end{array}%
\right) ,
\end{equation}%
and the time-reversal operation $\mathcal{T}i\mathcal{T}^{-1}=-i$. Then, the
Hamiltonian in the momentum space satisfies $\left( \mathcal{PT}\right)
H_{k}\left( \mathcal{PT}\right) ^{-1}=H_{k}$. In addition, chiral symmetry
exists when $J=0$ or the present synthetic magnetic flux is $\Phi =n\pi +\pi
/2$ ($n\in \mathbb{Z}$), where $Je^{\pm i\Phi }=0$ or $\left( -1\right)
^{n}\left( \pm iJ\right) $. Here, the chiral operator is defined as
\begin{equation}
\mathcal{C=}\left(
\begin{array}{ccc}
0 & 0 & 1 \\
0 & -1 & 0 \\
1 & 0 & 0%
\end{array}%
\right) ,
\end{equation}%
and the Hamiltonian in the momentum space satisfies $\mathcal{C}H_{k}%
\mathcal{C}^{-1}=-H_{k}$. In other conditions, chiral symmetry is absent. $%
H_{k}$ can possess a nonzero flat band through destructive interference
without the confinement of chiral symmetry. $H_k$ has three
energy bands, which can be analytically obtained through solving a cubic
equation
\begin{equation}
E_k^{3}-\left( 2s_{k}^{2}+J^{2}-\gamma ^{2}\right) E_k-2s_{k}^{2}J\cos \Phi
=0,  \label{EqEk}
\end{equation}
where $s_{k}=v+r\cos k$. The couplings $v$, $J$, and $r$ expand the energy
bands; the energy bands shrink in the presence of gain and loss.

\section{Flat band}

\label{III} A flat band is nondispersive and independent of the momentum $k$%
. Destructive interference is crucial for the formation of a flat band,
which can induce decoupling and isolation. To investigate the flat band in
the non-Hermitian lattice, we start from a Hermitian situation of $J=\gamma
=0$ to elucidate the formation of a flat band in a quasi-1D lattice. When $%
J=\gamma =0$, the system reduces to a cross-stitch lattice, as schematically
illustrated in Fig.~\ref{fig1}(b). If the cross-stitch coupling is zero ($%
r=0 $), the quasi-1D lattice possesses three flat bands because all the unit
cells are disconnected. The isolated trimer in each unit cell has three
eigen energies $0$ and $\pm \sqrt{2}v$, where the eigenstate of the $j$-th
unit cell for zero eigen energy is $(A_{j},B_{j},C_{j})=(1/\sqrt{2},0,-1/%
\sqrt{2})$; the vanishing distribution probability of the zero energy
eigenstate on site $B_{j}$ indicates that antisymmetric excitation in sites $%
A_{j}$ and $C_{j}$ destructively interferes at site $B_{j}$.

The cross-stitch coupling connects the nearest neighbor sites in the upper ($%
A$) and lower ($C$) sublattices with the central sublattice ($B$). In the
presence of cross-stitch coupling ($r\neq 0$), antisymmetric excitation in
sites $A_{j}$ and $C_{j}$ destructively interferes at sites $B_{j\pm 1}$; in
addition, antisymmetric excitation in the unit cells $j\pm 1$ destructively
interferes at site $B_{j}$. Thus, the central sublattice is effectively
decoupled with a vanishing distribution probability on the entire sublattice
$B$, and a zero-energy flat band is maintained in the presence of
cross-stitch coupling.

Consider the non-Hermitian lattice with $J,\gamma \neq 0$ in Fig. \ref{fig1}%
(a). The analysis of cross-stitch lattice indicates that a flat band is
formed when the sublattices $A$ and $C$ destructively interfere at the
central sublattice $B$. In the following, we calculate the
condition for maintaining such a destructive interference; that is the
condition for the existence of a flat band in the non-Hermitian lattice. By
acting $H_{k}$ [Eq. (\ref{Hk})] on $f_{k}=[1,0,-1]^{\mathrm{T}}/\sqrt{2}$, a
Schr\"{o}dinger equation $H_{k}f_{k}=E_{k}f_{k}$ is yielded, which gives
\begin{equation}
H_{k}\left(
\begin{array}{c}
1 \\
0 \\
-1%
\end{array}%
\right) =\left(
\begin{array}{c}
i\gamma -Je^{i\Phi } \\
0 \\
i\gamma +Je^{-i\Phi }%
\end{array}%
\right) =E_{k}\left(
\begin{array}{c}
1 \\
0 \\
-1%
\end{array}%
\right),
\end{equation}%
where $f_{k}=[1,0,-1]^{\mathrm{T}}/\sqrt{2}$ is an eigen state of $H_{k}$
that requires
\begin{equation}
i\gamma -Je^{i\Phi }=-\left( i\gamma +Je^{-i\Phi }\right) ,
\label{Constrain}
\end{equation}%
which gives $\gamma =J\sin \Phi $. Substituting $\gamma =J\sin \Phi $ back
into the Schr\"{o}dinger equation $H_{k}f_{k}=E_{k}f_{k}$, we obtain the
eigen energy $E_{k}=-J\cos \Phi $; which is independent of the momentum $k$.
Therefore, a flat band is formed under the condition
\begin{equation}
\gamma _{\mathrm{FB}}=J\sin \Phi ,  \label{gamma}
\end{equation}%
the flat band is induced by the interplay of non-Hermiticity and synthetic
magnetic flux. Correspondingly, the eigen energy of the flat band is
\begin{equation}
E_{\mathrm{FB}}=-J\cos \Phi .  \label{EFB}
\end{equation}%
Notably, the flat band energy is entirely real, and it is tunable through
different matches between synthetic magnetic flux and non-Hermiticity,
instead of pining to zero energy in the Hermitian limit of the cross-stitch
lattice at $J=\gamma =0$.

The condition given in Eq.~(\ref{gamma}) for the formation of the flat band
can be alternatively obtained as follows. The cross-stitch coupling of a
lattice causes the dispersion in the energy bands; effectively decoupling
sublattice $B$ under destructive interference enables the isolation of the
unit cells, thereby forming the flat band. In the flat band, the triangular
lattice in each unit cell effectively reduces into a $\mathcal{PT}$%
-symmetric dimer with asymmetric coupling between sublattices $A$ and $C$ in
the form of 
\begin{equation}
H_{\mathcal{PT}\text{\textrm{-dimer}}}=\left(
\begin{array}{cc}
i\gamma & Je^{i\Phi } \\
Je^{-i\Phi } & -i\gamma%
\end{array}%
\right) .
\end{equation}%
The asymmetric coupling phase factor $e^{\pm i\Phi }$ changes the
eigenstates without varying the pair of eigen energies $\pm \sqrt{%
J^{2}-\gamma ^{2}}$. The destructive interference at the central sublattice $%
B $ requires antisymmetric excitation on sublattices $A$ and $C$, where we
obtain identical constrain of Eq.~(\ref{Constrain}) by acting $H_{\mathcal{PT%
}\text{\textrm{-dimer}}}$ on the antisymmetric state $[1,-1]^{\mathrm{T}}/%
\sqrt{2}$. Consequently, antisymmetric excitation is one of the two
eigenstates of the $\mathcal{PT}$-symmetric dimer, and the corresponding
eigen energy is $E_{\mathrm{FB}}$. Antisymmetric excitation is a cage
solution confined in the single unit cell~\cite{CEC}.

Any antisymmetric excitation of each unit cell is confined in the single
unit cell; thus, the eigen functions of the flat band can be expressed as
\begin{equation}
\left\vert \psi _{\mathrm{FB}}\right\rangle =\left( 2\Omega \right)
^{-1/2}\sum_{j=1}^{N}\zeta _{j}(a_{j}^{\dagger }-c_{j}^{\dagger })\left\vert
\mathrm{vac}\right\rangle .
\end{equation}%
The renormalization coefficient is $\Omega =\sum_{j=1}^{N}\left\vert \zeta
_{j}\right\vert ^{2}$ and $\zeta _{j}$ is an arbitrary number. Antisymmetric
excitation is diffractionless with a constant intensity~\cite{RamezaniFB},
different from that in a non-Hermitian flat band entirely constituted by EPs~%
\cite{LeykamFB,LGePR}.

\begin{figure}[thb]
\includegraphics[ bb=0 0 305 210, width=8.8 cm, clip]{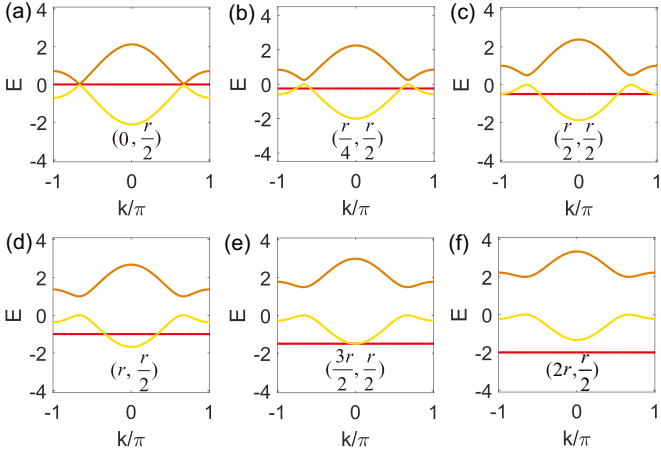}
\caption{Energy band structures with the flat band at different $(J\left\vert \cos \Phi \right\vert ,v)$ values (marked at the bottom). All
the energy bands are entirely real. The degree of non-Hermiticity is $\protect\gamma =J\sin {\Phi }$, $\Phi =\protect\pi /3$, and $r=1$. $k=-\protect\pi $ and $k=\protect\pi $ represent an identical
point.} \label{fig2}
\end{figure}

When $J=0$ or $\Phi =n\pi +\pi /2$ ($n\in \mathbb{Z}$), $H_{k}$ is chiral
symmetric ($\mathcal{C}H_{k}\mathcal{C}^{-1}=-H_{k}$) and supports a zero
energy flat band; the eigen function of $H_{k}$ for $\Phi =\pi /2$ is given
by%
\begin{equation}
f_{k}=[1,i\left( J-\gamma \right) /(v+r\cos k),-1]^{\mathrm{T}}.
\end{equation}%
The flat band states are CLSs~\cite{SFlachPRB95}, as will be discussed in
the next section. In a Hermitian lattice ($\gamma =0$), the flat band
appears in the absence of asymmetric coupling; $E_{\mathrm{FB}}=0$ when $J=0$
[Fig.~\ref{fig2}(a)] or $E_{\mathrm{FB}}=\left( -1\right) ^{n+1}J$ when $%
\Phi =n\pi $ ($n\in
\mathbb{Z}
$). The bands experience a conical intersection for $J=0$, similar to that
in the Lieb lattice~\cite{Julku,LeykamAPX16}.

\begin{figure}[b]
\includegraphics[ bb=0 0 160 160, width=7.0 cm, clip]{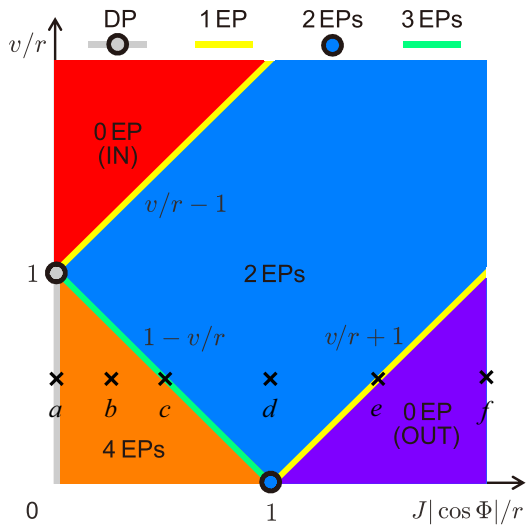}
\caption{Phase diagram of the flat band for $\Phi \neq n\pi +\pi /2$ ($n\in \mathbb{Z}
)$. The energy bands for the system at parameters marked by points $a$-$f$ are depicted in Figs.~\ref{fig2}(a)-\ref{fig2}(f), respectively.}
\label{figPD}
\end{figure}

\begin{figure*}[tb]
\includegraphics[ bb=0 0 540 310, width=16.8 cm, clip]{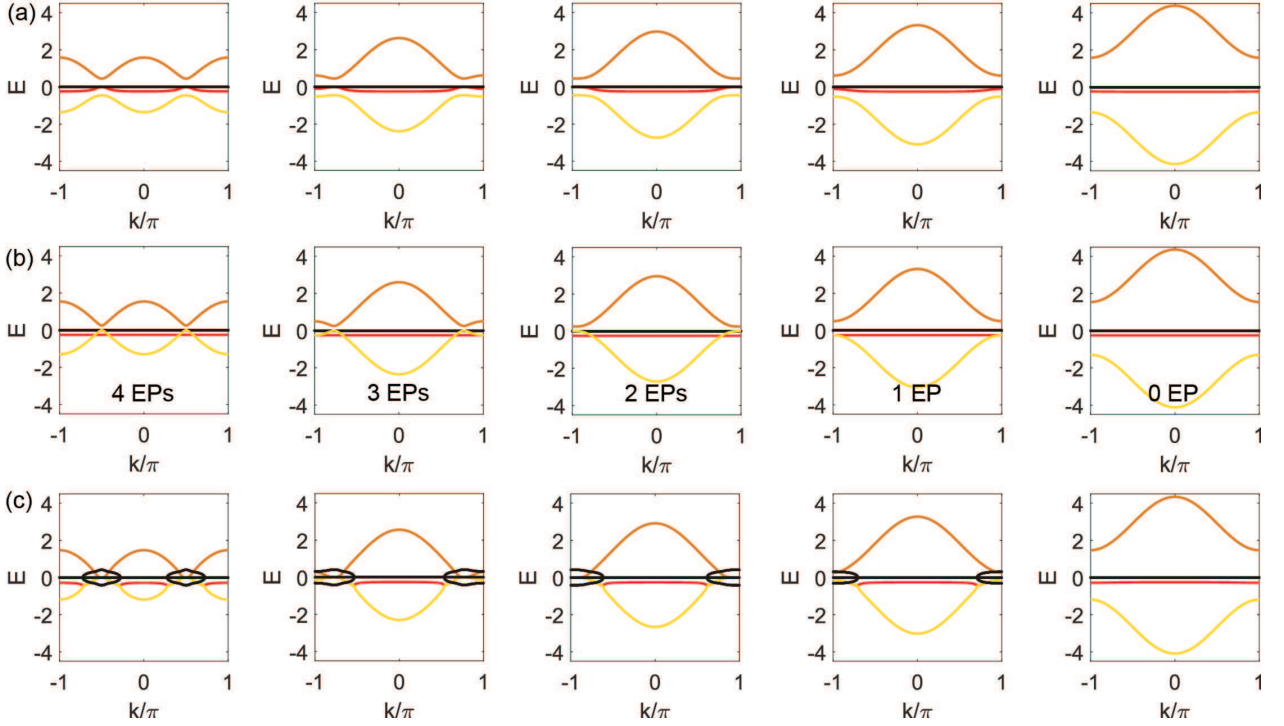}
\caption{Complex band structures at various $v$ and $\protect\gamma$ values
for $J=1/2$, $\Phi=\protect\pi/3$, and $r=1$. (a) $\protect\gamma=\protect%
\gamma_{\mathrm{FB}}/2$, (b) $\protect\gamma=\protect\gamma_{\mathrm{FB}}$,
(c) $\protect\gamma=3\protect\gamma_{\mathrm{FB}}/2$ for $v=0$, $3/4$, $1$, $%
5/4$, and $2$ from the left to the right panels, respectively. As $v$ increases in (b), the flat band energy is unchanged, but the
dispersive bands expand; thus, the band intersections reduce from four EPs
to zero EP. $k=-\protect\pi $ and $k=\protect\pi $ represent an identical
point.}
\label{fig3}
\end{figure*}

Figure~\ref{fig2} depicts the energy bands of the system at
various intracell couplings when the flat band is formed. The intracell
couplings expand the energy bands; the flat band energy depends on the
intracell coupling $J$, but is independent of the intracell coupling $v$
[Eq.~(\ref{EFB})]. Thus, the flat band appears in the band gap of the
dispersive bands at a strong coupling $v$, and the flat band is the lowest
energy band outside the dispersive bands at a strong coupling $J$. The
intercell cross-stitch coupling $r=0$ is a trivial case for the existence
of flat band; thus, we choose $r$ as the unit to demonstrate the influence
of the intracell couplings $v$ and $J$ on the energy bands. Figure~\ref%
{figPD} is the phase diagram of the flat band in the parameter space of intracell couplings, where the phases are classified by the
number of EPs in the flat band embedded in the continuum. The
boundaries $v/r-1$ and $1-v/r$ indicate that the band touching occurs at $%
k=\pi$; while the boundary $v/r+1$ indicates that the band touching occurs at $k=0$.
Typical cases of the energy bands are depicted in Fig.~\ref{fig2}, which
reflects the relation between the flat band and dispersive bands; the
representative energy bands in the red region of Fig.~\ref{figPD} is
depicted in Fig.~\ref{fig3}(b) for $v=2$ ($0$ EP).

For $J=0$, the lattice system is chiral symmetric. A zero-energy flat band
appears, the other two dispersive bands are symmetric, and $%
E_{k}[E_{k}^{2}-2(v+r\cos k)^{2}+\gamma ^{2}]=0$. The intracell coupling $v$
generates three energy bands, and the band gaps are proportional to this
coupling; the intercell coupling $r$ induces dispersion of the upper and
lower energy bands. The dispersive bands are gapped when the intercell
coupling is weak, that is, $v/r>1$. A dispersive band each locates above and
below the zero-energy flat band (belongs to the red region of
Fig.~\ref{figPD}). At a strong intercell coupling $v/r\leqslant 1$, the
coupling $v+r\cos k$ in $H_{k}$ can vanish; thus, the band gaps vanish and
two dispersive bands touch at $\left\vert k_{\mathrm{DP}}\right\vert
=\arccos \left( -v/r\right) $ in the Brillouin zone (the gray
line in Fig.~\ref{figPD}). The band touching points are diabolic points of
two-fold Hermitian degeneracy, which move from $\pi /2$ toward $\pi $ as the
intracell coupling $v$ increases from $0$ to $r$. The energy bands tighten
under the influence of gain and loss \cite{Segev2011}, this can
be seen in Fig.~\ref{fig3} by comparing Figs.~\ref{fig3}(a),~\ref{fig3}(b),
and~\ref{fig3}(c). Without the band gap ($v/r\leqslant 1$),
the energy bands are vulnerable to non-Hermiticity; any nonzero $\gamma $
brings the system into a $\mathcal{PT}$-symmetry broken phase and the eigen
energy starts to become complex around $k_{\mathrm{DP}}$. The band gaps at $%
v/r>1$ protect the system by rendering it robust against a certain degree of
non-Hermiticity $\gamma =\sqrt{2}(v-r)$, where the band gaps vanish at $%
\left\vert k_{\mathrm{EP}}\right\vert =\pi $. In the presence of
non-Hermiticity, the gapless band touching points are EPs, where both the
real and imaginary parts of the spectrum are gapless.

For $J\neq 0$, the flat band energy shifts from zero and intersects a
dispersive band at the EPs
\begin{equation}
\left( v+r\cos k_{\mathrm{EP}}\right) ^{2}=J^{2}\cos ^{2}\Phi .
\label{EPCondition}
\end{equation}
This condition is useful for understanding the phase diagram
(Fig.~\ref{figPD}) and is obtained as follows. If the system supports a flat
band, Eq.~(\ref{EqEk}) reduces to $E_{k}^{3}-(2s_{k}^{2}+ J^{2}\cos
^{2}\Phi) E_{k}-2s_{k}^{2}J\cos \Phi =0$. At the band touching point, two
among the three roots of $E_k$ are equal; thus, $(2s_{k}^{2}+J^{2}\cos
^{2}\Phi) ^{3}/3^{3}-( s_{k}^{2}J\cos \Phi ) ^{2}=0$ should be satisfied.
After simplification, we obtain $(8s_{k}^{2}+J^{2}\cos ^{2}\Phi)
(s_{k}^{2}-J^{2}\cos ^{2}\Phi) ^{2}=0$; then, the condition Eq.~(\ref%
{EPCondition}) for the band intersection EPs is acquired. The number of EPs
is determined from the competitions between the coupling strengths as shown
in Fig.~\ref{figPD}. At weak intracell couplings $v$, $J$, four EPs are
obtained from $v+r\cos k_{\mathrm{EP}}=\pm J \cos \Phi$; at moderate
intracell couplings $v$, $J$, two EPs are obtained from $v+r\cos k_{\mathrm{%
EP}}=+J \cos \Phi$ ($v+r\cos k_{\mathrm{EP}}=-J \cos \Phi$); and EP
disappears in the situation that either the intracell coupling $v$ or $J$ is
very strong [$k_{\mathrm{EP}}$ has no real solution in Eq.~(\ref{EPCondition})], which corresponds to the case that the flat band appears isolatedly inside
or outside the dispersive bands with zero EP, respectively.

In the orange and cyan regions of Fig.~\ref{figPD}, the band gap diminishes
as the degree of non-Hermiticity increases and the flat band appears when
the band gap closes at $\gamma =\gamma _{\mathrm{FB}}$~\cite{RamezaniFB}.
Therefore, the flat band intersects the dispersive band and becomes the
bound states in the continuum~\cite{CTChanBIC,Hsu}. $\gamma _{\mathrm{FB}}$
is the $\mathcal{PT}$-symmetric phase transition point of the non-Hermitian
lattice. In the red (violet) region of Fig.~\ref{figPD}, the flat band at $%
\gamma =\gamma _{\mathrm{FB}}$ is isolated inside (outside) the dispersive
bands with band gaps, where EP does not exist. The {boundary between the
gapped red (violet) phase and the gapless blue phase with two EPs is }$v-r$ (%
$v+r$), where two EPs merge to one at the edges (center) of the Brillouin
zone at $\left\vert k_{\mathrm{EP}}\right\vert =\pi $ ($0$). The boundaries
of distinct phases in the phase diagram indicate that the band gap vanishes
for stronger critical non-Hermiticity at $\gamma _{\mathrm{c,-}}$ ($\gamma _{%
\mathrm{c,+}}$) with
\begin{equation}
\gamma _{\mathrm{c,\pm }}=\sqrt{2\left( v\pm r\right) ^{2}+J^{2}-3\sqrt[3]{%
\left( v\pm r\right) ^{4}J^{2}\cos ^{2}\Phi }}.
\end{equation}%
where $\gamma _{\mathrm{c,\pm }}$ is obtained from the cubic equation of the
energy bands of $H_{k}$, that is, $\mathrm{det}\left( H_{k}-E_{k}I_{3\times
3}\right) =0$. The energy bands of $H_{k}$ are no longer entirely real when
the band gap vanishes. We can directly verify that $\gamma _{\mathrm{c,\pm }%
}>\gamma _{\mathrm{FB}}$ in the phases where the energy bands are gapped.

The intracell couplings $J$ and $v$ expand the energy bands and the band gap
widths, and the intercell coupling $r$ induces dispersion. Thus, the
relation between energy bands varies with the variation in the competition
of system couplings. The flat band may appear inside [Fig.~\ref%
{fig3}(b) ($0$ EP)], at the intersection [Figs.~\ref{fig2}(a)-\ref{fig2}(e)], or outside [Fig.~\ref{fig2}(f)] the dispersive
bands. At a strong intracell coupling, $J\left\vert \cos \Phi \right\vert
>v+r $, the flat band appears outside the dispersive bands [Fig.~\ref{fig2}(f)]. At a moderate intracell coupling, $\left\vert
v-r\right\vert <J\left\vert \cos \Phi \right\vert <v+r$, the flat band
intersects the lower dispersive band and two EP2s (two-state coalescence)
appear at $k_{\mathrm{EP}}$ [Fig.~\ref{fig2}(d)]. When $%
J\left\vert \cos \Phi \right\vert <r-v$, the flat band intersects the lower
dispersive band and four EP2s appear at $k_{\mathrm{EP}}$ [Fig.~\ref{fig2}%
(b)]; moreover, with no band gap at $\gamma =0$ [Fig.~\ref{fig2}(a)], the
non-Hermiticity brings two dispersive bands closer and the two band touching
points split into four EP2s. When EP2s appear at the center ($k_{\mathrm{EP}%
}=0$) and the edges ($k_{\mathrm{EP}}=\pm \pi $) of the Brillouin zone,
pairs of EP2s merge and leave three EP2s [Fig.~\ref{fig2}(c)],
two EP2s [Fig.~\ref{fig2}(d)], and one EP2 [Fig.~\ref{fig2}(e)]. At a weak intracell coupling, $J\left\vert \cos \Phi
\right\vert <v-r$, the dispersive bands are gapped and the flat band appears
inside [Fig.~\ref{fig3}(b) ($0$ EP)].

When the system is under chiral symmetry at $J=0$ or $\Phi =n\pi +\pi /2$ ($%
n\in
\mathbb{Z}
$), the EPs in the flat band are EP3 (three-state coalescence). Two EP3s $%
\left\vert k_{\mathrm{EP}}\right\vert =\arccos \left( -v/r\right) $ exist in
the flat band for $v<r$, they merge to one EP3 at $\left\vert k\right\vert
=\pi $ for $v=r$, and EP3 disappears for $v>r$. All the EPs can be
determined from $\Delta _{k}=27s_{k}^{4}J^{2}\cos ^{2}\Phi
-[(2s_{k}^{2}+J^{2}-\gamma ^{2}]^{3}=0$.

The flat band is partially flat except for $\gamma =\gamma _{\mathrm{FB}}$,
and the flatness is lost near $k_{\mathrm{EP}}$ and increases as $\gamma $
approaches $\gamma _{\mathrm{FB}}$. When $\gamma <\gamma _{\mathrm{FB}}$,
the lattice system is in the exact $\mathcal{PT}$-symmetric phase, and the
energy bands become closer as non-Hermiticity increases. The flat band
formed at the $\mathcal{PT}$-symmetric phase transition point $\gamma
=\gamma _{\mathrm{FB}}$, is entirely flat with the bands touching and band
gaps closing. When $\gamma >\gamma _{\mathrm{FB}}$, the system is in the
broken $\mathcal{PT}$-symmetric phase, the flatness is most robust around
the center of the Brillouin zone and is severely lost at the edges of the
Brillouin zone. In Fig.~\ref{fig3}, the energy bands of $H_{k}$ are depicted
for parameters $J=1/2$, $\Phi =\pi /3$, and $r=1$ at different $v$ and $%
\gamma $ values. At a weak non-Hermiticity $\gamma <\gamma _{%
\mathrm{FB}}$, the energy bands are gapped [Fig.~\ref{fig3}(a)]; at an
appropriate non-Hermiticity $\gamma =\gamma _{\mathrm{FB}}$, the flat band
appears [Fig.~\ref{fig3}(b)]; at a strong non-Hermiticity $\gamma >\gamma _{%
\mathrm{FB}}$, the system is in the broken $\mathcal{PT}$-symmetric phase~%
\cite{PTRev} and the energy bands are complex [Fig.~\ref{fig3}(c)]. Figures~%
\ref{fig3}(a)-\ref{fig3}(c) show the energy bands of $H_{k}$ for $\gamma
=\gamma _{\mathrm{FB}}/2$, $\gamma _{\mathrm{FB}}$, and $3\gamma _{\mathrm{FB%
}}/2$, respectively at $v$ varying from $0$ to $2$. The bands expand as the
intracell coupling $v$ increases and shrink as $\gamma $ increases. At a weak intracell coupling, $v<r-J|\cos \Phi |$, where four EPs
appear at critical $\gamma _{\mathrm{FB}}$, the two lower bands form two
windows separated by the four EPs at large non-Hermiticity and the central
band near $\left\vert k\right\vert =0,\pi $ is nearly flat; at a moderate
intracell coupling, $|r-J|\cos \Phi ||<v<r+J|\cos \Phi |$, where two EPs
appear at critical $\gamma _{\mathrm{FB}}$, the partial flat band near $%
\left\vert k\right\vert =\pi $ vanishes and the central band near $k=0$ is
nearly flat, whose flatness increases as $v$ increases. The flat band
appears in the band gap between the dispersive bands for a strong intracell
coupling $v>r+J|\cos \Phi |$, where EP disappears. The bifurcation of the
energy bands at the EPs significantly destroys the band flatness. A nearly
flat region appears away from the EPs, shrinks, and finally vanishes with
increasing non-Hermiticity; subsequently, $H_{k}$ has one real band and two
conjugate pure imaginary bands.

\section{Compact localized states}

\label{IV} The flat band energy is insensitive to the boundary conditions.
At $\Phi =n\pi +\pi /2$ ($n\in
\mathbb{Z}
$), the lattice is chiral symmetric and has a zero-energy flat band. The
flat band CLSs are confined in three unit cells; only the edge modes are
confined in two unit cells under an open boundary condition. The
non-Hermitian lattice has two pairs of edge modes that are localized at two
boundaries of the lattice and $N-4$ confined modes that are localized inside
the lattice. They are formed through destructive interference at the central
sublattice $B$.

The eigen functions of the upper and lower sublattices $A$ and $C$ should
satisfy $\psi _{A_{j}}+\psi _{C_{j}}=0$. The subscript in the wave functions
$\psi $ indicates the site number. Note that sites $A_{j}$ and $C_{j}$ have
an identical coupling $v$ to $B_{j}$ in the $j$-th unit cell and identical
couplings $r/2$ to $B_{j\pm 1}$ in the neighbor unit cells; thus, the
contributions from $A_{j}$ and $C_{j}$ cancel each other and vanish in the
Schr\"{o}dinger equations of $B_{j}$ and $B_{j\pm 1}$. Consequently, the
steady-state Schr\"{o}dinger equations for the sublattice $B$ are satisfied
because $\psi _{A_{j}}+\psi _{C_{j}}=0$ and due to the zero-energy of the
flat band, that is, ${0}=${$E_{\mathrm{FB}}\psi _{B_{j}}=v(\psi
_{A_{j}}+\psi _{C_{j}})+\left( r/2\right) (\psi _{A_{j-1}}+\psi
_{C_{j-1}})+\left( r/2\right) (\psi _{A_{j+1}}+\psi _{C_{j+1}})$. }

\begin{figure}[tb]
\includegraphics[ bb=0 0 360 230, width=8.8 cm, clip]{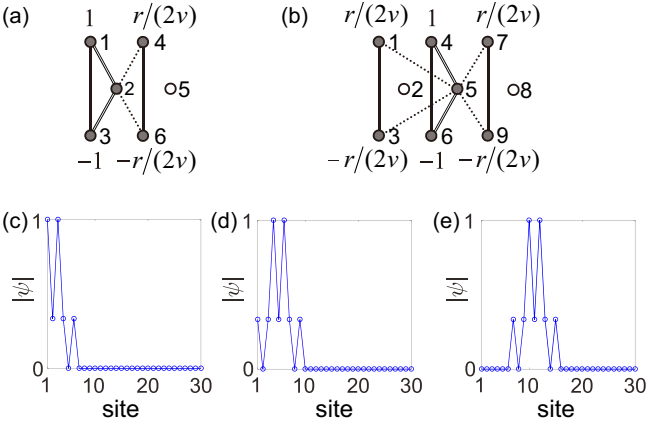}
\caption{Schematic of destructive interference for the lattices consisting of (a) two and (b) three unit cells under an open boundary condition.
The couplings connected to the unoccupied passive sublattice (hollow circles) are hidden. The wave function of the central site ${B_j}$ is $i(J-\gamma)/v$; $A_{j}$ and $C_{j}$ are marked at the top and bottom of the schematic. The occupied sites are represented by gray circles. (c, d) Edge mode; (e) inner confined mode. The parameters are $\Phi=\pi/2$, $J=1$, $r=1$, $v=3/2$, and $\gamma=1/2$.}
\label{fig4}
\end{figure}

If the eigen function $\psi _{B_{j}}=0$, $B_{j}$ and $A_{j+1}$, $C_{j+1}$
are effectively decoupled; moreover, the interference of $A_{j}$ and $C_{j}$
is destructive at $B_{j+1}$. Therefore, $B_{j}=0$ results in a decoupling
between the $j$-th and $(j+1)$-th unit cells, forming a state that is
localized on the left side of the $j$-th unit cell (similarly, a state that
is localized on the right side of the $j$-th unit cell is formed with the
decoupling of $B_{j}$ and $A_{j-1}$, $C_{j-1}$). Specifically, $\psi
_{B_{1}}=0$ is possible when $\gamma =J$, which is the critical
non-Hermiticity {$\gamma _{\mathrm{FB}}$} for $\Phi =n\pi +\pi /2$ ($n\in
\mathbb{Z}
$). This leads to the decoupling of the whole sublattice $B$ from the
lattice system $H$ and the eigen function being antisymmetric and
independent of the lattice size. $\psi _{B_{2}}=0$ corresponds to a state
confined in the first two unit cells. The Schr\"{o}dinger equations for the
lattice system consisting of two unit cells under an open boundary condition
are%
\begin{eqnarray}
i\dot{\psi}_{A_{1}} &=&i\gamma \psi _{A_{1}}+v\psi _{B_{1}}+Je^{i\Phi }\psi
_{C_{1}}+\frac{r}{2}\psi _{B_{2}}, \\
i\dot{\psi}_{B_{1}} &=&v\left( \psi _{A_{1}}+\psi _{C_{1}}\right) +\frac{r}{2%
}\left( \psi _{A_{2}}+\psi _{C_{2}}\right) , \\
i\dot{\psi}_{C_{1}} &=&-i\gamma \psi _{C_{1}}+v\psi _{B_{1}}+Je^{-i\Phi
}\psi _{A_{1}}+\frac{r}{2}\psi _{B_{2}}, \\
i\dot{\psi}_{A_{2}} &=&i\gamma \psi _{A_{2}}+v\psi _{B_{2}}+Je^{i\Phi }\psi
_{C_{2}}+\frac{r}{2}\psi _{B_{1}}, \\
i\dot{\psi}_{B_{2}} &=&v\left( \psi _{A_{2}}+\psi _{C_{2}}\right) +\frac{r}{2%
}\left( \psi _{A_{1}}+\psi _{C_{1}}\right) , \\
i\dot{\psi}_{C_{2}} &=&-i\gamma \psi _{C_{2}}+v\psi _{B_{2}}+Je^{-i\Phi
}\psi _{A_{2}}+\frac{r}{2}\psi _{B_{1}}.
\end{eqnarray}%
The non-normalized zero mode eigen function is $[\varphi ,\phi ]^{\mathrm{T}%
} $ as illustrated in Fig.~\ref{fig4}(a), where $\varphi \equiv \left( \psi
_{A_{1}},\psi _{B_{1}},\psi _{C_{1}}\right) =(1,iJ/v-i\gamma /v,-1)$ and $%
\phi \equiv \left( \psi _{A_{2}},\psi _{B_{2}},\psi _{C_{2}}\right)
=[r/\left( 2v\right) ,0,-r/\left( 2v\right) ]$.

The energy band structure shown in Fig.~\ref{fig4}(a) only appears at the
lattice boundary, $[\varphi ,\phi ]^{\mathrm{T}}$, which is the zero mode
for a lattice with more unit cells connected at the right side of the
two-unit-cell lattice because $\psi _{B_{2}}=0$ (i.e., the zero mode of the
extended lattice is unoccupied on the additional unit cells). The
structure's mirror reflection corresponds to the zero mode for a lattice
with its left side capable of connecting more unit cells. $\psi
_{B_{1}}=\psi _{B_{3}}=0$ corresponds to a state that is confined in the
first three unit cells $[\phi ,\varphi ,\phi ]^{\mathrm{T}}$, as illustrated
in Fig.~\ref{fig4}(b). The structure shown in Fig.~\ref{fig4}(b) can be
formed inside the lattice $H$ with more than three unit cells; $[\phi
,\varphi ,\phi ]^{\mathrm{T}}$ characterizes the probability distributions
of the zero mode for a lattice with additional unit cells connected at both
the left and right sides.

Under an open boundary condition, $H$ has two pairs of edge modes $\psi _{%
\mathrm{EMi}},\mathcal{PT}\psi _{\mathrm{EMi}}$ in the form
\begin{equation}
\psi _{\mathrm{EMi}}=\left[ \varphi ,\phi ,\cdots \right] ^{\mathrm{T}},
\label{EMi}
\end{equation}%
and $\psi _{\mathrm{EMii}},\mathcal{PT}\psi _{\mathrm{EMii}}$ in the form
\begin{equation}
\psi _{\mathrm{EMii}}=\left[ \phi ,\varphi ,\phi ,\cdots \right] ^{\mathrm{T}%
}.  \label{EMii}
\end{equation}%
The edge modes only distribute in the two or three unit cells at the lattice
boundaries; $\cdots $ in Eqs.~(\ref{EMi})-(\ref{EMii}) represents the unit
cells $\left( A_{j},B_{j},C_{j}\right) $ with vanishing occupation$\ \left(
0,0,0\right) $. The edge modes in Eqs.~(\ref{EMi})-(\ref{EMii}) are depicted
in Figs.~\ref{fig4}(c) and~\ref{fig4}(d), respectively.

The inner lattice CLSs are in the form
\begin{equation}
\psi _{\mathrm{CLS}}=\left[ \cdots ,\phi ,\varphi ,\phi ,\cdots \right] ^{%
\mathrm{T}},  \label{CM}
\end{equation}%
which is independent of the boundary conditions. The confined modes
distribute in the three unit cells that are localized inside the lattice
with the passive sublattice $B$ at the edges of the three unit cells being
unoccupied. The $\mathcal{PT}$-symmetric counterpart $\mathcal{PT}\psi _{%
\mathrm{CLS}}$ belongs to the confined modes. The confined modes for the
unoccupied first two and last five unit cells are depicted in Fig.~\ref{fig4}%
(e). All zero modes are degenerate, and their superpositions are also zero
modes. The trapping mechanism is destructive interference assisted by the
synthetic magnetic flux inside the unit cells~\cite{JC,CEC,SLOL}. Each pair
of edge modes [Eqs.~(\ref{EMi})-(\ref{EMii})] can form one pair of $\mathcal{%
PT}$-symmetric zero modes; the confined modes [Eq.~(\ref{CM})] can form
another $N-4$ $\mathcal{PT}$-symmetric zero mode.

\begin{figure}[tb]
\includegraphics[ bb=0 0 285 110, width=8.8 cm, clip]{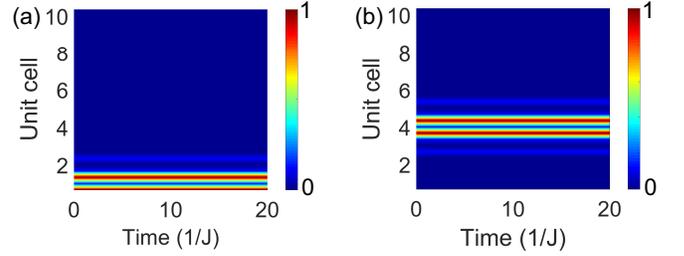}
\caption{Time evolution of initial excitations. The intensity $|\psi (t)|^2$
is depicted. The parameters are $\Phi=\pi/2$, $J=1$, $r=1$, $v=3/2$, and
$\gamma=1/2$, which are identical to the parameters in Fig.~\ref{fig4}.} %
\label{fig5}
\end{figure}

Any zero mode excitation is confined inside that lattice without escaping
and its dynamics exhibits no diffraction effect. This is numerically
simulated in a lattice of size $N=30$ with $10$ unit cells. The dynamics for
the excitation of the zero edge mode confined in two unit cells [Fig.~\ref%
{fig4}(c)] is depicted in Fig.~\ref{fig5}(a). The zero mode is confined at
the lattice boundary without spreading or escaping. The dynamics of the
excitation of an inner lattice that confines the zero mode [Fig.~\ref{fig4}%
(e)] is depicted in Fig.~\ref{fig5}(b). The excitation of any superposition
of CLSs is localized and diffractionless in the dynamical process. This
dynamics is considerably different from that of a confined polynomial
increase of excitation in the non-Hermitian lattices, where flat bands are
entirely constituted by EPs \cite{LeykamFB,LGePR}.

Equations (\ref{EMi})-(\ref{CM}) are valid for the zero-energy flat band
when $J=0$. For the lattice without chiral symmetry, that is, $\Phi \neq
n\pi +\pi /2$ ($n\in
\mathbb{Z}
$), the eigen function of the central sublattice $B$ is zero for the flat
band $E=E_{\mathrm{FB}}$ at $\gamma =\gamma _{\mathrm{FB}}$ under both
periodical and open boundary conditions. The flat band is constituted by the
CLSs that are confined in the single unit cell. Its eigen functions are
given by Eqs.~(\ref{EMi})-(\ref{CM}) with $\varphi $ replaced by $\varphi
=\left( 1,0,-1\right) $ and $\phi $ unchanged; the eigen functions change
into the superposition of antisymmetric excitations in the neighboring
single unit cells.

\section{Summary}

\label{V} In this work, we first report a novel configuration
that supports a flat band due to the destructive interference at an
appropriate match between the synthetic magnetic flux and non-Hermiticity.
The flat band energy is flexible at different appropriate matches instead of
pinning to zero, and the flat band can intersect the dispersive bands or
appear isolatedly inside/outside the dispersive band gap; the flat band states
form bound states in the continuum when energy bands intersect.

A quasi-1D $\mathcal{PT}$-symmetric non-Hermitian lattice is proposed, whose
unit cell is a triangular lattice enclosed synthetic magnetic flux. The
synthetic magnetic flux is attributed to the AB-type nonreciprocal phase
factor in the intracell coupling. We demonstrated the flat band induced by
the interplay of synthetic magnetic flux and non-Hermiticity. The phase
diagram characterizes the relation between the flat band and dispersive
bands. When the flat band intersects the dispersive band, the flat band
appears at the $\mathcal{PT}$-symmetric phase transition point. The eigen
functions of the flat band are compact localized states, which are confined
within a few unit cells inside or at the edges of the non-Hermitian lattice.

Our findings provide insights into the interplay of synthetic magnetic flux
and non-Hermiticity in the application of controllable flat band and bound
states in the continuum in non-Hermitian metamaterials. Further studies of
the Aharonov-Bohm cage, nonreciprocal localization, and anomalous edge modes
would be interesting in both theoretical and experimental aspects.

\section*{ACKNOWLEDGMENTS}

This work was supported by NSFC (Grant No. 11605094).

\end{document}